\documentstyle[12pt,psfig]{article}

\newcommand{\noi}{\noindent}
\def\ifmath#1{\relax\ifmmode #1\else $#1$\fi}
\def\half{\ifmath{{\textstyle{1 \over 2}}}}

\def\3quarter{{\textstyle{3 \over 4}}}

\def\ra{\rightarrow}

\def\lf{\leaders\hbox to 1em{\hss.\hss}\hfill}
\def\21{$SU(2) \ot U(1)$}
\def\ne{\hbox{$\nu_e$ }}
\def\nm{\hbox{$\nu_\mu$ }}
\def\nt{\hbox{$\nu_\tau$ }}

\def\O{\hbox{$\cal O$ }}
\def\L{\hbox{$\cal L$ }}

\def\mnt{\hbox{$m_{\nu_\tau}$ }}

\def\neu{\hbox{neutrino }}

\def\eq#1{{eq. (\ref{#1})}}
\def\Eq#1{{Eq. (\ref{#1})}}

\def\fig#1{{Fig. (\ref{#1})}}

\def\VEV#1{\left\langle #1\right\rangle}
\let\vev\VEV

\def\lsim{\raise0.3ex\hbox{$\;<$\kern-0.75em\raise-1.1ex\hbox{$\sim\;$}}}
\def\gsim{\raise0.3ex\hbox{$\;>$\kern-0.75em\raise-1.1ex\hbox{$\sim\;$}}}
\def\half{{1\over 2}}
\def\beq{\begin{equation}}
\def\eeq{\end{equation}}
\def\bef{\begin{figure}}
\def\eef{\end{figure}}
\def\bet{\begin{table}}
\def\eet{\end{table}}
\def\bea{\begin{eqnarray}}
\def\ba{\begin{array}}
\def\ea{\end{array}}
\def\bi{\begin{itemize}}
\def\ei{\end{itemize}}
\def\ben{\begin{enumerate}}
\def\een{\end{enumerate}}
\def\ra{\rightarrow}
\def\ot{\otimes}

\def\eea{\end{eqnarray}}
\def\apj#1#2#3{          {\it Astrophys. J. }{\bf #1} (19#2) #3}

\def\nat#1#2#3{          {\it Nature }{\bf #1} (19#2) #3}
 
\def\np#1#2#3{           {\it Nucl. Phys. }{\bf #1} (19#2) #3}
\def\pl#1#2#3{           {\it Phys. Lett. }{\bf #1} (19#2) #3}
\def\pr#1#2#3{           {\it Phys. Rev. }{\bf #1} (19#2) #3}

\def\prl#1#2#3{          {\it Phys. Rev. Lett. }{\bf #1} (19#2) #3}

\def\n.c.#1#2#3{         {\it Nuovo Cim. }{\bf #1} (19#2) #3}
\def\r.n.c.#1#2#3{       {\it Riv. del Nuovo Cim. }{\bf #1} (19#2) #3}

\def\ppnp#1#2#3{           {\it Prog. Part. Nucl. Phys. }{\bf #1} (19#2) #3}

\def\ijmp#1#2#3{           {\it Int. J. Mod. Phys. }{\bf #1} (19#2) #3}

\def\ip{in preparation}
\begin{document}
\thispagestyle{empty}
\begin{titlepage}
\begin{center}
\hfill FTUV/96-57\\
\hfill IFIC/96-65\\
\hfill TAC/96-022\\
\vskip 0.2cm
{\Large \bf Primordial Nucleosynthesis, Majorons and Heavy Tau Neutrinos \\}
\vskip 0.8cm
 A.D. Dolgov$^1$ 
\footnote{Also: ITEP, 113259, Moscow, Russia.}, 
S.~Pastor$^2$, J. C. Rom\~ao$^3$
\footnote{E-mail fromao@alfa.ist.utl.pt}
and J. W. F. Valle$^2$
\footnote{E-mail: valle@flamenco.ific.uv.es}\\
{\sl $^1$ Theoretical Astrophysics Center;
                              Juliane Maries Vej 30\\
                              DK-2100 K\o benhavn, Denmark}\\
{\sl $^2$Instituto de F\'{\i}sica Corpuscular - C.S.I.C.\\
Departament de F\'{\i}sica Te\`orica, Universitat de Val\`encia\\
46100 Burjassot, Val\`encia, Spain\\
 http://neutrinos.uv.es}\\
\vskip .2cm
{\sl $^3$ Inst. Superior T\'ecnico, Dept. de F\'{\i}sica \\
Av. Rovisco Pais, 1 - 1096 Lisboa, Codex, Portugal}\\
\vskip .2cm
\newcommand{\ttbs}{\char'134}
\newcommand{\AmS}{{\protect\the\textfont2
  A\kern-.1667em\lower.5ex\hbox{M}\kern-.125emS}}
\hyphenation{author another created financial paper re-commend-ed
nucleo-syn-thesis}
\vskip 1cm
 
\begin{abstract}

We determine the restrictions imposed by primordial nucleosynthesis
upon a heavy tau neutrino, in the presence of \nt annihilations into 
Majorons, as expected in a wide class of particle physics models of 
neutrino mass. We determine the equivalent number of light 
neutrino species $N_{eq}$ as a function of \mnt and the 
\nt-\nt-Majoron coupling $g$.
We show that for theoretically plausible $g$ values $\gsim 10^ {-4}$ 
present nucleosynthesis observations can not rule out \nt masses in the
MeV range. Moreover, these models give $N_{eq} \leq 3$ in the \nt mass 
region 1-10 MeV, for very reasonable values of $g \geq 3 \times 10^ {-4}$.
The evasion of the cosmological limits brings new interest to the
improvement of the present laboratory limit on the \nt mass which can
be achieved at a tau-charm factory.

\end{abstract}
\end{center}
\vfill
\end{titlepage}
\newpage
\section{Introduction}

Despite great experimental efforts, the tau-neutrino still remains 
as the only one which can have mass in the MeV range. The present
experimental limit on its mass is \cite{eps95}: 
\beq
m_{\nu_\tau} <  23 \, \mbox{MeV}
\label{mlab}
\eeq
Further progress will have to wait  for the improvements expected
at future tau-charm or B factories \cite{jj}. On the other
hand,  many particle physics models of massive neutrinos
lead to a tau neutrino with mass in the MeV range \cite{beyond}.
Moreover such a neutrino may have  interesting cosmological 
implications \cite{ma1}. It is therefore interesting to examine 
critically the cosmological constraints. 

The first comes from the critical density argument \cite{boundCosm}.
However, as has been widely illustrated with many particle physics 
models where  neutrinos acquire their mass by the spontaneous
violation of a global lepton number symmetry \cite{fae}, this
limit can be avoided due to the existence of fast \nt decays
\cite{V,mu,774} and/or annihilations \cite{GR,mu} into 
 Majorons. Although the Majoron was first introduced in the
context of the seesaw model \cite{CMP} the spontaneous breaking
of lepton number can be realized in many different models. There
is only one important constraint on its properties following from
the precision measurements of the invisible Z width at LEP,
namely the Majoron must be mostly singlet under the \21 symmetry.
It has been noted that, in many models of this type the relic
\nt number density can be  depleted well below the required value
for all masses obeying \eq{mlab}. 

In order to demonstrate the cosmological viability of the MeV
tau neutrino we must also consider the restrictions that follow
from primordial nucleosynthesis considerations \cite{sarkar}.
In the standard model, these  
rule out $\nu_\tau$ masses in the range \cite{ckst,dr}: 
\beq
0.5\, \mbox{MeV} <  m_{\nu_\tau} <  35\, \mbox{MeV}
\label{mns}
\eeq
This would imply that $m_{\nu_\tau} < 0.5 $ MeV is the
 nucleosynthesis limit for the case of a Majorana 
tau neutrino. Here we will only assume that \nt is a
Majorana particle,  which is the most likely possibility. This assumes
 for the maximum allowed effective number of extra 
neutrino species $\Delta N_{eq}$ during nucleosynthesis
either 0.4 or 0.6. Recent contradictory data on the
primordial deuterium abundance \cite{dhigh,dlow} may cast some 
doubts on the validity of this assumption (for recent analysis see 
refs. \cite{cst,olive}). In particular, if $\Delta N_{eq} = 1$ is allowed
\cite{olive}, there may be an open window for neutrino mass somewhere 
near 20 MeV. However it has been shown in ref. \cite{noneq}  that 
this window actually does not exist, when one carefully takes into 
account the influence of non-equilibrium electronic neutrinos on the 
neutron-to-proton ratio. These neutrinos would come from massive 
$\nu_\tau$ annihilations $ \nu_\tau \nu_\tau \rightarrow \nu_e \nu_e$. 

However one knows that new  interactions capable of depleting
MeV \nt density in the cosmic plasma are needed, at some level, 
in order to comply with the limit on the relic neutrino density. 
It is therefore reasonable to analyse their possible 
effect in relation with the primordial nucleosynthesis
constraints \cite{steigman}. 

\begin{figure}
\centerline{\protect\hbox{\psfig{file=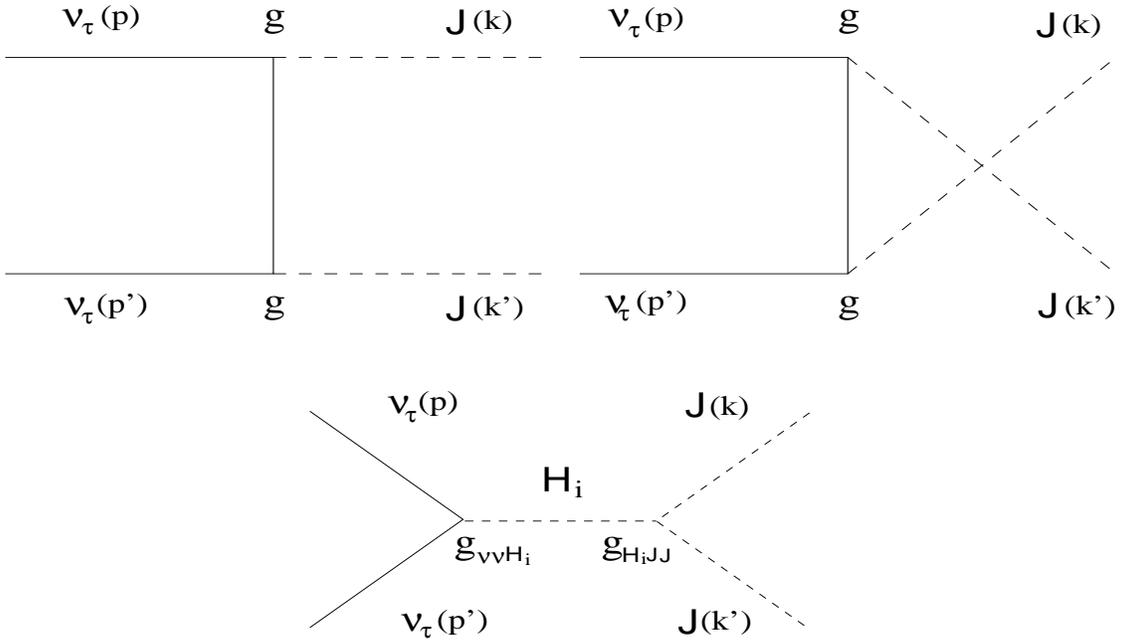,height=8.5cm,width=0.9\textwidth}}}
\caption{Feynman diagrams for annihilations  of tau neutrinos
into Majorons.} 
\label{fig2}
\end{figure}

In this paper we  analyse the effect of neutrinos with large 
annihilation cross sections into Majorons. 
In order to compute the relevant  annihilation rates  we must 
parametrize the majoron interactions. 
These arise from the diagrams shown in \fig{fig2}.
The t-channel diagram is present in all Majoron models,
while the strength of the s-channel scalar exchange diagram 
is somewhat model-dependent. 

One way of writing the couplings of Majorons to neutrinos is 
using the fact that the Majorons are Nambu-Goldstone bosons 
and hence have derivative couplings. This is the so called {\it
polar coordinate} method. The other method is to use a pseudoscalar
interaction, sometimes refered to as the {\it cartesian method}. 
The two methods are equivalent, even for second order processes 
as we are considering here, if we include {\it all} the Feynman 
diagrams contributing at that order to the process of interest
\nt \nt $\ra$ J J
\footnote{
Although equivalent, for models with a large number of 
scalars and where the Majoron is a linear combination of the
imaginary parts of several fields, like the model of Ref. 
\cite{MASIpot3}, the cartesian method is more convenient.}. 
In our calculations  throughout this paper  we will use the 
cartesian method of parametrizing the majoron interactions.
Though we must in principle include also the s-channel diagram 
in \fig{fig2}, we will neglect this contribution. We explicitly 
show in the Appendix, that it is justified in our case to use only 
the t-channel contribution in order to derive a {\it conservative} 
limit on neutrino mass $m_{\nu_\tau}$ and majoron coupling $g$. 

We have determined the restrictions imposed by primordial 
nucleosynthesis upon such a heavy tau neutrino in the presence
of \nt annihilations into Majorons. We show that if the \nt \nt 
Majoron coupling constant exceeds $g \gsim 10^ {-4}$ or so, 
a large \nt mass in the MeV range is allowed by the present
upper bounds on the extra number of neutrino species.
As a result one cannot rule out any values of the \nt mass up 
the present laboratory limit of \eq{mlab}. 
\footnote{In fact, with a larger coupling constant $g \gsim 
10^ {-3}$ it may be possible for a stable MeV \nt to obey the
critical density limit, suggesting a possible role of \nt as 
dark matter.}.

We also show how such $g$ values are theoretically plausible
in the context of the most attractive elementary particle  physics 
models where MeV tau neutrinos arise, and which are based upon the
spontaneous violation of lepton number. 

\section{Evolution of $\nu_\tau$ number density in the presence
 of \nt \newline
annihilations}

Massive tau neutrinos certainly interact with leptons via the 
standard weak interactions, $\nu_\tau \nu_\tau \leftrightarrow 
\nu_{e,\mu}\bar{\nu}_{e,\mu}$, $e^+ e^-$, as assumed in refs. 
\cite{ckst,dr}. Moreover, in many  particle physics where neutrinos 
acquire mass from the spontaneous violation of a global lepton 
number symmetry \cite{fae} heavy neutrinos, such as the \nt,
annihilate to Majorons \nt \nt $\ra$ J J via the diagonal coupling
\begin{equation}
\label{ann}
\L =  i \half g J \nu_\tau^T \sigma_2 \nu_\tau \: + \: H.c.
\end{equation}
where $\nu_\tau$ represents a two-component Majorana spinor, in
the notation of ref. \cite{774,2227,BFD}. 

This corresponds, in the usual four-component notation to
\beq
\L = i \half g J \overline{\nu}_{\tau} \gamma_5 \nu_{\tau} 
\label{lagint}
\eeq
The corresponding elastic processes do not change particle 
densities, but as long as they are effective they maintain 
all species with the same temperature.

We now comment on the cosmological bound provided
by the critical density argument \cite{boundCosm}.
In order to be consistent with cosmological limits, the 
relic abundance of the heavy Majorana tau neutrinos must 
be suppressed over and above what is provided by the standard 
model charged and neutral current weak interactions, as well
as those derived from \fig{fig2}.
This happens automatically in many Majoron models, where neutrinos 
decay with lifetimes shorter than required by the critical density 
constraint \cite{fae,V,mu,774}. For example, in Majoron models of the 
seesaw-type a massive $\nu_\tau$ will typically decay with lifetimes 
shorter than the one required in order to obey the critical density 
bound, but longer than the relevant nucleosynthesis time, as 
illustrated in figure 18 of ref. \cite{fae}. Another example is 
provided by the model of ref. \cite{MASIpot3}. A \nt lifetime 
estimate was given for this model in Fig. 1 of ref. \cite{RPMSW}. 
It is seen explicitly that a \nt of mass in the MeV range of 
interest to us is expected to be stable on the nucleosynthesis 
time scale, but decays with lifetimes shorter than required by the 
critical density bound. This corresponds to a range of off-diagonal
neutrino-majoron couplings $10^{-10} > g_{off-diagonal} > 10^{-13}$,
which  naturally occurs in many models.

For simplicity, we will assume from now in this paper that the 
massive $\nu_\tau$'s decay with lifetimes shorter than required 
by the critical density bound, but are stable on the  time scale
relevant for nucleosynthesis considerations. The more general
case where both decays and annihilations are simultaneously active 
on the nucleosynthesis time scale will be treated elsewhere 
\cite{dprv2}.

\subsection{Before Weak Decoupling}
\label{bef}

Let us assume first that all species are interacting so
that they have the same temperature. The evolution of the 
$\nu_\tau$ density can be found from the corresponding Boltzmann equation,
\begin{equation}
\dot{n}_{\nu_\tau} + 3Hn_{\nu_\tau} = -\sum_{i=J,e,\nu_{e,\mu}}
\VEV{ \sigma_i v}
\left(n_{\nu_\tau}^2-(n_{\nu_\tau}^{eq})^2
\frac{n_i^2}{(n_i^{eq})^2}\right)
\label{Boltz1}
\end{equation}
In this expression $\VEV{ \sigma_i v}$ is the thermal
average of the annihilation cross section times the \nt relative 
velocity $v$. Using the convention for the momenta as in  figure
\ref{fig2}, its value for the process 
$\nu_\tau \nu_\tau' \leftrightarrow x_i x_i'$ is
\footnote{Here $v=[(pp')^2-m_{\nu_\tau}^4]^{1/2}/E_pE_{p'}$.}
\begin{eqnarray}
\VEV{ \sigma_i v} \equiv \frac{1}{(n_{\nu_\tau}^{eq})^2}
\int d\Pi_{\nu_\tau} d\Pi_{\nu_\tau'} d\Pi_{x_i} d\Pi_{x_i'}~
(2\pi)^4 \delta^4 (p+p'-k-k') \nonumber \\
\times \mid M \mid ^2 e^{-E_p/T} e^{-E_{p'}/T}
\label{sigmav}
\end{eqnarray}
Here we have assumed  kinetic equilibrium amongst the different 
species, as well as Boltzmann statistics. By   $\mid M \mid ^2$ 
we denote  the invariant amplitude obtained with the usual
Feynman rules for Majorana neutrinos \cite{774,2227,BFD},
 summed over all spins (and averaged over initial spins).
Moreover we have set $d\Pi_A \equiv d^3p_A/(2\pi)^32E_{p_A}$.

Following reference \cite{GonGel} we express $\VEV{ \sigma_iv }$ 
as a single integral using the dimension-less variable
$x \equiv m_{\nu_\tau}/T$, 
\begin{equation}
\label{intsigmav1}
\VEV{ \sigma_iv } = \frac{x}{8m_{\nu_\tau}^5 K_2^2(x)} 
\int_{4m_{\nu_\tau}^2}^{\infty}ds~(s-4m_{\nu_\tau}^2)\sigma_i(s)
\sqrt{s} K_1\Bigl (\frac{x\sqrt{s}}{m_{\nu_\tau}}\Bigr )
\end{equation}
where $K_i(x)$ are the modified Bessel functions of order $i$
(see for instance \cite{tablas}) and $s=(p+p')^2$ is the invariant
of the process $\nu_\tau \nu_\tau' \leftrightarrow x_i x_i'$.
Using the new variable $\eta \equiv 1-4m_{\nu_\tau}^2/s$ instead of $s$,
\begin{equation}
\label{intsigmav2}
\VEV{ \sigma_iv } = \frac{4x}{K_2^2(x)} 
\int_{0}^{1}d\eta~\frac{\eta}{(1-\eta)^{7/2}}
\sigma_i(\eta) K_1\Bigl (\frac{2x}{\sqrt{1-\eta}}\Bigr )
\end{equation}
The cross-sections of the different annihilation processes
are listed below. For annihilations to Majorons we have
\footnote{The general formula is given in the Appendix, \eq{sigmaJJ}.}
\begin{equation}
\label{sigmaJ}
\sigma_J(\eta) = \frac{g^4}{128\pi} \frac{1-\eta}{m_{\nu_\tau}^2 \eta}
\Bigl [\ln \Bigl (\frac{1+\sqrt{\eta}}{1-\sqrt{\eta}}\Bigr ) -
2\sqrt{\eta}\Bigr ]~.
\end{equation}
where we have divided by 2! in order to account for identical
Majorons in the final state and  divided by 4 in order to account
the \nt spin factors. For the standard weak interaction-induced
annihilations $\nu_\tau \bar{\nu}_\tau \leftrightarrow 
f_i \bar{f}_i$, in the limit of {\sl massless} products we take
\begin{equation}
\label{sigmaiw}
\sigma_i(\eta) = \frac{2G_F^2}{3\pi} \frac{m_{\nu_\tau}^2 \sqrt{\eta}}
{1-\eta} (b_{Li}^2 + b_{Ri}^2)~,
\end{equation}
where $b_L^2 + b_R^2 = 1/2$ for $i=\nu_{e,\mu}$ and
$b_L^2 + b_R^2 = 2((-1/2+\sin^2\theta_W)^2+(\sin^2\theta_W)^2) \simeq
0.25$ for $i=e$.

One may write evolution equations analogous to \eq{Boltz1} 
for the other species present in the plasma, namely $\nu_{e,\mu}$ and $e^\pm$.
However we assume that the weak and electromagnetic 
interactions are effective enough to keep $\nu_{e,\mu}$'s and $e$'s 
densities in their equilibrium values, $n_k = n_k^{eq}$ for 
$k = \nu_{e,\mu},e$. Thus we are left with a system of just 
two coupled Boltzmann equations:
\begin{eqnarray}
\dot{n}_{\nu_\tau} + 3Hn_{\nu_\tau} = -\sum_{i=e,\nu_{e,\mu}} 
\VEV{\sigma_i v}
(n_{\nu_\tau}^2-(n_{\nu_\tau}^{eq})^2) 
-\VEV{\sigma_J v} \left(n_{\nu_\tau}^2- (n_{\nu_\tau}^{eq})^2
\frac{n_J^2}{(n_J^{eq})^2} \right) \equiv S_{\nu_\tau}
\label{Boltz2}
\end{eqnarray}
\begin{equation}
\dot{n}_J + 3Hn_J = \VEV{\sigma_J v}
(n_{\nu_\tau}^2-(n_{\nu_\tau}^{eq})^2
\frac{n_J^2}{(n_J^{eq})^2}) \equiv S_J
\label{Boltz3}
\end{equation}

Now let us briefly describe our calculations. First we normalized
the number densities to the number density of a massless neutrino
species, $n_0 \simeq 0.181T^3$, introducing the quantities
$r_\alpha \equiv n_\alpha/n_0$, where $\alpha = \nu_\tau,J$, and
the corresponding equilibrium functions $r_\alpha^{eq}$. We then
have for the time derivative of $n_\alpha$
$$
\dot{n}_\alpha = \dot{T} \frac{dn_\alpha}{dT} =
S_\alpha - 3Hn_\alpha
$$
$$
\frac{dn_\alpha}{dT} = n_0\frac{dr_\alpha}{dT} +
r_\alpha\frac{3}{T} n_0
$$
or, equivalently,
\begin{equation}
\label{drdt}
\frac{dr_\alpha}{dT} = \Bigl (\frac{S_\alpha}{n_0} - 3Hr_\alpha
\Bigr )\frac{1}{\dot{T}} - \frac{3}{T}r_\alpha
\end{equation}
On the other hand, the time derivative of the temperature is obtained
from covariant energy conservation law
\begin{equation}
\label{einstein}
\dot{\rho} = -3H(\rho+P)
\qquad
\rightarrow
\qquad
\dot{T} = -3H(\rho+P)\frac{1}{d\rho/dT}
\end{equation}
where $\rho$ is the total energy density and $P$ is the pressure. Finally,
as $\rho = \rho (T,r_J,r_{\nu_\tau})$ we can rewrite
$$
\frac{d\rho}{dT} = \frac{\partial \rho}{\partial T} +
\frac{\partial \rho}{\partial r_J} \frac{dr_J}{dT} +
\frac{\partial \rho}{\partial r_{\nu_\tau}} \frac{dr_{\nu_\tau}}{dT}~,
$$
and for the normalized particle densities one has
\begin{eqnarray}
\label{evol1}
\frac{dr_{\nu_\tau}}{dT} = - \Sigma_{\nu_\tau}\Bigl (
\frac{\partial \rho}{\partial T} +
\frac{\partial \rho}{\partial r_J} \frac{dr_J}{dT} +
\frac{\partial \rho}{\partial r_{\nu_\tau}}
\frac{dr_{\nu_\tau}}{dT}\Bigr ) -\frac{3}{T}r_{\nu_\tau} \\
\label{evol2}
\frac{dr_J}{dT} = - \Sigma_J\Bigl (
\frac{\partial \rho}{\partial T} +
\frac{\partial \rho}{\partial r_J} \frac{dr_J}{dT} +
\frac{\partial \rho}{\partial r_{\nu_\tau}}
\frac{dr_{\nu_\tau}}{dT}\Bigr ) -\frac{3}{T}r_J
\end{eqnarray}
where, for $\alpha = \nu_\tau,J$, we have introduced
\begin{equation}
\label{defSigma}
\Sigma_\alpha \equiv \frac{1}{\rho+P}\Bigl (
\frac{S_\alpha}{3Hn_0} - r_\alpha \Bigr )
\end{equation}
The final Boltzmann system for the normalized particle densities 
is obtained from \eq{evol1} and \eq{evol2} introducing the 
dimension-less variable $x$ previously
defined. Denoting $r' \equiv dr/dx$, we have
\begin{eqnarray}
\label{evol3}
r'_{\nu_\tau}\Bigl (1+\Sigma_{\nu_\tau}\frac{\partial \rho}
{\partial r_{\nu_\tau}}\Bigr ) + r'_J \Sigma_{\nu_\tau}
\frac{\partial \rho}{\partial r_J} = 
\Sigma_{\nu_\tau} \frac{T}{x} \frac{\partial \rho}{\partial T} +
\frac{3}{x} r_{\nu_\tau} \\
\label{evol4}
r'_J\Bigl (1+\Sigma_J\frac{\partial \rho}
{\partial r_J}\Bigr ) + r'_{\nu_\tau} \Sigma_J
\frac{\partial \rho}{\partial r_{\nu_\tau}} = 
\Sigma_J \frac{T}{x} \frac{\partial \rho}{\partial T} +
\frac{3}{x} r_J
\end{eqnarray}
This system is valid as long as the tau neutrinos are coupled
to the weak interactions. The following is the complete set of entries
in equations (\ref{evol3}) and (\ref{evol4}) for the equilibrium
quantities, total energy density and pressure, respectively:
\begin{eqnarray}
r_{\nu_\tau}^{eq} =\frac{1}{0.181\pi^2}x^3 I_1(x) \:,\:\: \: 
r_J^{eq} = \frac{2}{3}  \nonumber \\
\rho = \rho_{\nu_0}+\rho_e+\rho_\gamma+\rho_J+\rho_{\nu_\tau} =
\frac{3\pi^2}{10}T^4 \left(1+\frac{1}{6}r_J+0.06x \frac{I_2(x)}{I_1(x)}
r_{\nu_\tau}\right) \nonumber \\
\label{totalP}
P = P_{\nu_0}+P_e+P_\gamma+P_J+P_{\nu_\tau} =
\frac{\pi^2}{10}T^4 \left(1+\frac{1}{6}r_J+0.06x \frac{I_3(x)}{I_1(x)}
r_{\nu_\tau}\right)~.
\end{eqnarray}
In these expressions we have introduced the integral functions
$I_j$, where $j=1,2,3$, defined as
\begin{eqnarray}
I_1(x) = \int_{0}^{\infty}du~u^2 \exp(-x\sqrt{1+u^2}) \nonumber \\
I_2(x) = \int_{0}^{\infty}du~u^2 \sqrt{1+u^2}
\exp(-x\sqrt{1+u^2}) \nonumber \\
I_3(x) = \int_{0}^{\infty}du~\frac{u^4}{\sqrt{1+u^2}}
\exp(-x\sqrt{1+u^2})
\label{dedIs}
\end{eqnarray}

\subsection{Past Weak Decoupling}

Once the $\nu_\tau$'s decouple from the standard weak interactions, 
they remain in contact only with Majorons. Then one has two different
plasmas, one formed by $\nu_\tau$'s and $J$'s and the other by the
rest of particles, each one with its own
temperature\footnote{Eventually the massless neutrinos will also
decouple from the second plasma, while the $e^+e^-$ pairs will
annihilate to photons, thus generating the well known
$T_{\nu_0}-T_\gamma$ difference.} (denoted
as $T$ and $T_\gamma$). Let us define now the variables
$$
x = \frac{m_{\nu_\tau}}{T} \:,
\qquad
y = \frac{m_{\nu_\tau}}{T_\gamma}
$$
We assume that the photon temperature evolves in the
usual way, $\dot{y} = Hy$. The evolution equation of the $\nu_\tau$ and
$J$ number densities are now simplified versions of (\ref{Boltz2})
and (\ref{Boltz3}), because $S_{\nu_\tau} \equiv -S_J$,
\begin{eqnarray}
\dot{n}_{\nu_\tau} + 3Hn_{\nu_\tau} = -S_J \nonumber \\
\dot{n}_J + 3Hn_J = S_J 
\label{Boltz5}
\end{eqnarray}
or, in terms of $r_\alpha$'s,
\begin{eqnarray}
r_{\nu_\tau}'= -\frac{S_J}{n_0Hy}\frac{dy}{dx} \nonumber \\
r_J'= -r_{\nu_\tau}'
\label{evol5}
\end{eqnarray}
Due to the second equation,  the Boltzmann system  reduces
 to a single evolution equation say, for $r_{\nu_\tau}$.
 However, one must still determine $dy/dx$ which
differs from unity because $T \neq T_\gamma$. An equation relating 
$y$ and $x$ is obtained
using the energy balance condition for the $\nu_\tau+J$ plasma. If
$\rho \equiv \rho_{\nu_\tau}+\rho_J$ and $P \equiv P_{\nu_\tau}+P_J$,
we can write
\begin{equation}
\label{Einstein2}
\dot{\rho} = -3H(\rho+P)~,
\qquad
\mbox{where} 
\quad H=\sqrt{\frac{8\pi \rho_{tot}}{3M_{pl}^2}}
\end{equation}
\begin{figure}
\centerline{\protect\hbox{\psfig{file=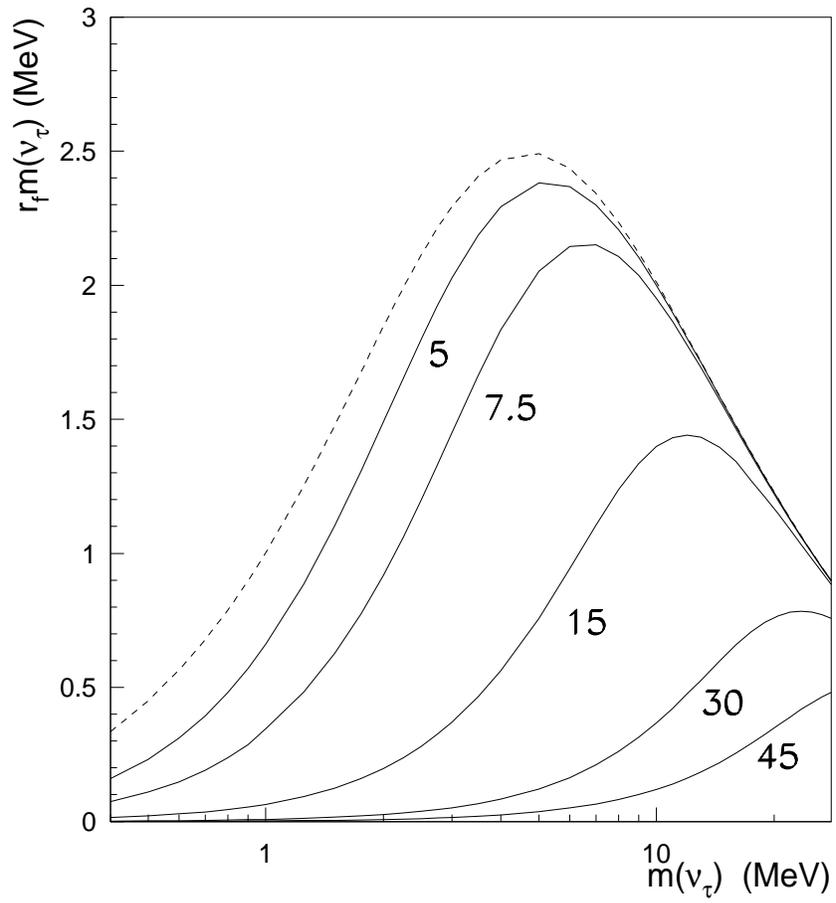,height=12cm,width=11cm}}}
\caption{Frozen values of $r_{\nu_\tau} m_{\nu_\tau}$ as a 
function of \mnt for the standard model ($g=0$) and for the 
Majoron model with different $g$ values in units of $10^{-5}$.}
\label{frozen}
\end{figure}
The expressions for $r_\alpha^{eq}$, $\rho$ and $P$ given
in equations \eq{totalP} need to be modified in order to 
take into account the fact that there are two distinct 
temperatures $T$ and $T_\gamma$. This leads to the 
following  equation
\begin{equation}
\label{dydx}
\frac{dy}{dx} = \frac{y \Bigl [\frac{\pi^2}{20} \frac{r_J}{x^2} +
\Bigl (\frac{I_4(x)}{I_1(x)} - \Bigl (\frac{I_2(x)}{I_1(x)}\Bigr )^2
\Bigr )r_{\nu_\tau}\Bigr ]}
{3\Bigl (0.06\frac{I_3(x)}{I_1(x)}r_{\nu_\tau} +
\frac{\pi^2}{60} \frac{r_J}{x}\Bigr ) -
\frac{r_{\nu_\tau}'}{H}\Bigl (\frac{\pi^2}{20x} -
0.18 \frac{I_2(x)}{I_1(x)}\Bigr )}~.
\end{equation}
Here we defined $I_4(x) \equiv -dI_2(x)/dx$.

In order to determine the final frozen density of \nt which will
be relevant during nucleosynthesis we have to solve numerically 
the corresponding set of differential equations. Before weak
decoupling these are (\ref{evol3}) and (\ref{evol4}), while 
after decoupling one should combine  \eq{evol5} and
\eq{dydx}, with the initial conditions $r_\alpha = r_\alpha^{eq}$,
$\alpha = J,\nu_\tau$ valid at high temperatures.

In \fig{frozen} we show the results of our calculations 
of the asymptotic (frozen) values of $r_{\nu_\tau} m_{\nu_\tau}$
as a function of \mnt for the standard model ($g=0$) and for
the Majoron model with different $g$ values.
Note that in the standard $g=0$ case we agree with the previous 
results of ref. \cite{dr} but get somewhat larger values than 
those obtained in ref. \cite{ckst}. We ascribe this small discrepancy
to the use, in ref.\cite{ckst} of an approximate expression for the
\nt energies, rather than the exact ones.

\section{Nucleosynthesis constraints on $(m_{\nu_\tau},g)$}

In this section we use the results obtained for the $\nu_\tau$
number density in order to constrain its mass from nucleosynthesis
arguments. The value of $r_{\nu_\tau}$ as a function of $(m_{\nu_\tau},g)$ 
is used in order to estimate the variation of the total energy density 
$\rho_{tot}=\rho_{R} + \rho_{\nu_\tau}$. In $\rho_{R}$ all relativistic 
species are taken into account, including Majorons and two massless 
neutrinos, whereas $\rho_{\nu_\tau}$ is the energy density of the 
massive $\nu_\tau$'s.
\begin{figure}
\centerline{\protect\hbox{\psfig{file=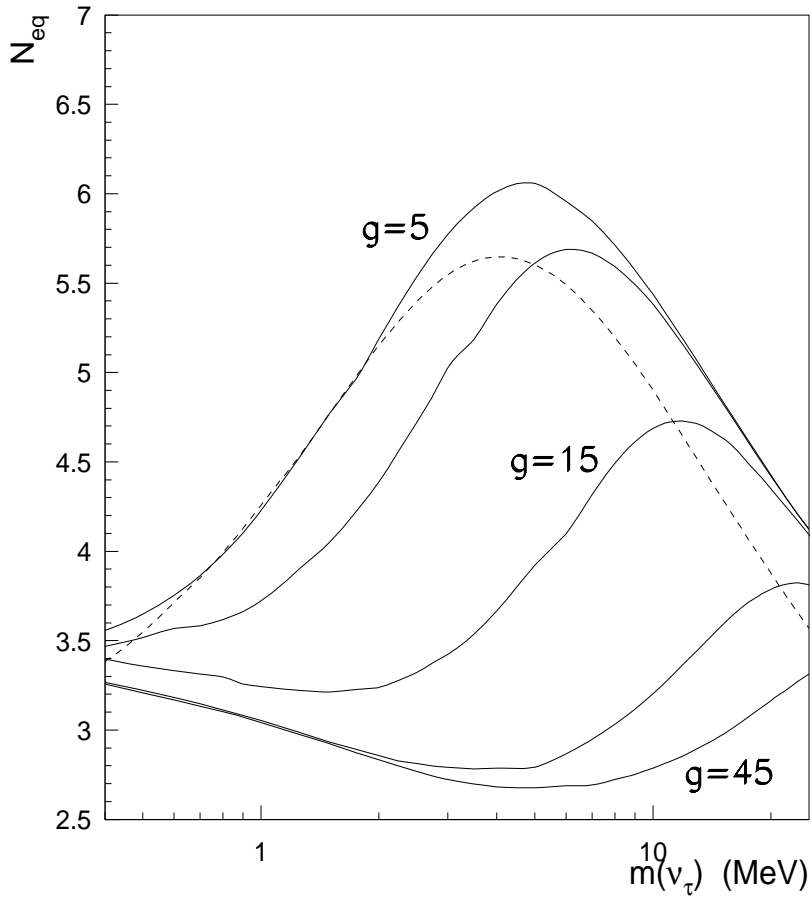,height=12cm,width=11cm}}}
\caption{Effective number of massless neutrinos equivalent to
the contribution of heavy \nt's  with different values of $g$ 
 in units of $10^{-5}$. For comparison, the dashed line
corresponds to the standard model case when $g=0$.}
\label{neffm}
\end{figure}

In order to compare with the standard model situation it is
convenient for us to express the effect of the \nt mass and that of the
presence of the Majoron in terms of an effective
 number of massless neutrino species ($N_{eq}$) which we 
calculate for each frozen value of $r_\tau(m_{\nu_\tau})$. 
In reality, the true value of  $r_\tau(m_{\nu_\tau})$ is
always larger than its frozen value, and we have taken this into
account in order to obtain reliable  limits in the low
\nt mass region.

 In order to derive the nucleosynthesis limits, first we 
developed a simple code for the numerical calculation of the 
neutron fraction $r_n$, as presented e.g. in ref. \cite{rnevol}, 
varying the value of $N_{eq}$. Then we incorporated 
$\rho_{tot}$ to this numerical code and performed the 
integration of the neutron-proton kinetic equations
for each pair of $(m_{\nu_\tau},g)$ values, where
$g$ is the  coupling constant which determines the 
strength of the \nt annihilation cross section.
Comparing $r_n(m_{\nu_\tau},g)$ with  $r_n(N_{eq})$
at $T_\gamma \simeq 0.075$ MeV (the moment when practically 
all neutrons are wound up in $^4$He), we can relate 
$(m_{\nu_\tau},g)$ to $N_{eq}$.

We repeated this calculation adapting Kawano's nucleosynthesis
code \cite{kawano} to the case of a massive tau neutrino, both 
in the standard  model and the Majoron extension. 
We have found that both methods are in good agreement.
The results for the numerical calculations of the equivalent 
number of massless neutrinos during nucleosynthesis with the use of 
Kawano's numerical code are shown in figure \ref{neffm}. 
For comparison the case of $g=0$ is shown (dashed line). 
{}From figure \ref{neffm} one can see that, in the asymptotic limit 
of very large \mnt the annihilation into Majorons is very inefficient 
(see \eq{sigmaJ}), so that the effective  $N_{eq}$ value is larger 
than in the standard $g=0$ case precisely by a factor $4/7$, which 
corresponds to the extra Majoron degree of freedom. Thus, if we 
take also $g$ very large we get just $N_{eq} = 2+4/7 \simeq 2.57$.
Of course this asymptotic limit is already experimentally
ruled out by the Aleph \nt mass limit \cite{eps95} and thus 
is not displayed. 
For \mnt values in the range from 10 to 23 MeV or so, $N_{eq}$ 
can be made acceptable, provided g is raised sufficiently. 
For the intermediate \nt mass region, 1-10 MeV, and $g> 3\times 10^{-4}$ 
the model may even give $N_{eq} \leq 3$, which is possibly supported by 
some of the observational data.

Finally, in the small \nt  mass limit the energy density of
\nt is roughly the same as that of the massless \ne or \nm, 
so that all $g$ values shown in the figure lead to the same asymptotic 
value $N_{eq} = 3+4/7 \simeq 3.57$, corresponding to the three
 massless neutrinos plus Majoron (instead of $2 + 4/7$ for a
very heavy \nt).
 However,  it might be that observations  eventually 
could lead to a tighter limit $N_{eq}^{max} \leq 3.57$. In such event a
simple way out is to have the Majoron out-of-equilibrium, 
which would require a very small $g$ value, $g < (2-3)\times 10^{-5}$,
so that the production of Majorons through annihilations of \nt's
would be negligible
\footnote{Of course such \mnt values are allowed
by nucleosynthesis in the absence of \nt annihilations.}.
Should the observations eventually lead to an even tighter limit
$N_{eq}^{max} \leq 3$ the situation is qualitatively different, as
it would raise a conflict with the standard model. A possible way 
to lower $N_{eq}$ below three provided by our model is to have a 
massive \nt in the MeV range and with a relatively strong coupling 
with Majorons. Indeed, one can see from Figure \ref{neffm} that, 
while it is not possible in the standard model to account for 
$N_{eq}^{max} \leq 3$, it is quite natural in our model to 
have $N_{eq}^{max} \leq 3$ for a wide range of intermediate 
tau neutrino masses and reasonable large values of the coupling 
constants $g$.

In summary, one sees that all \nt masses below 23 MeV are allowed by 
the nucleosynthesis condition $N_{eq} \leq N_{eq}^{max}$ if
$N_{eq}^{max} \geq 3.57$,  provided that the coupling between 
$\nu_\tau$'s and $J$'s exceeds a value of a few times $10^{-4}$. 
This situation seems at the moment compatible with the
experimental data, at least the $^4$He and $^7$Li 
determinations \cite{olive}.

\begin{figure}
\centerline{
\psfig{file=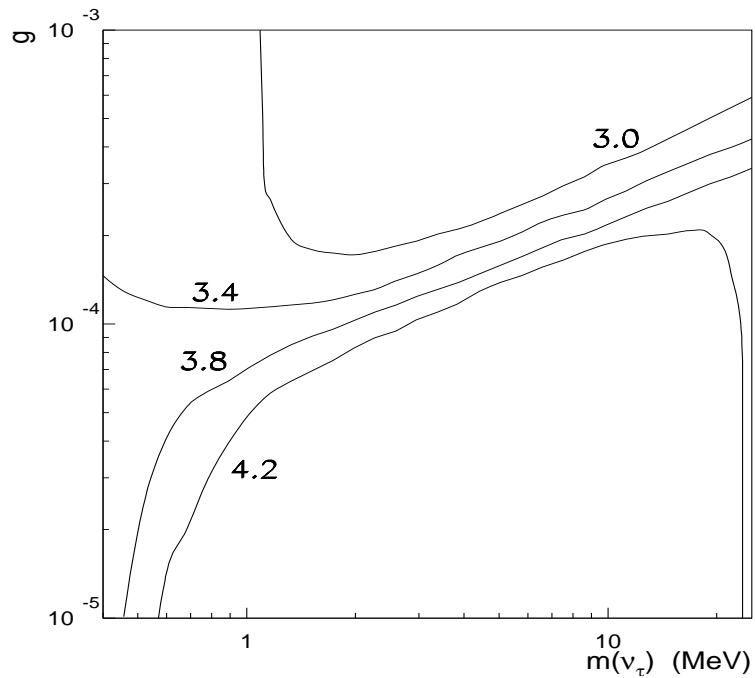,height=9cm,width=10cm}}
\caption{The values of $g(m_{\nu_\tau})$ above each line 
would be allowed by nucleosynthesis if one adopts the $N_{eq}^{max}
= 3, 3.4, 3.8, 4.2$ (from top to bottom). }
\label{neffmg}
\end{figure}

It is instructive to express the above results in the $m_{\nu_\tau}-g$ 
plane, as shown in figure \ref{neffmg}. The region above each curve is 
allowed for the corresponding $N_{eq}^{max}$.  

\section{Significance of the Nucleosynthesis Limits}

There has been a variety of Majoron models proposed in the
literature \cite{fae}. They are attractive extensions of the 
standard electroweak model where neutrinos acquire mass by
virtue of the spontaneous violation of a global lepton number symmetry. 
Apart from their phenomenological interest as extensions of the lepton
and/or Higgs sectors of the standard model \cite{beyond}, Majoron models 
offer the possibility of loosening the cosmological limits on neutrino 
masses, either because neutrinos decay or because they annihilate to 
Majorons. The first and most obvious of these is the limit that follows 
from the cosmological density argument \cite{V,mu}. As we saw in the 
previous section one can also place limits on a heavy tau neutrino 
with mass in the MeV range by using primordial element abundances. 
We have determined the restrictions imposed by primordial 
nucleosynthesis upon a heavy tau neutrino, in the presence of \nt \nt 
annihilations into Majorons. Our results are completely general and
may be compared to any bound characterized by an allowed value of 
$N_{eq}^{max}$. Given any $N_{eq}^{max}$ value one can readily
 obtain the allowed regions of \mnt and the Majoron coupling constant
$g$ as shown in \fig{neffmg}. As an example, a recent model-independent 
likelihood analysis of big bang nucleosynthesis based on $^4$He and $^7$Li 
determinations has claimed an upper limit  $N_{eq} < 4.0$ (at 95\% C.L.) 
\cite{olive}. From \fig{neffmg} this would imply that all \mnt masses 
are allowed, as long as $g$ exceeds $10^{-4}$ or so. However we believe 
that, in the present state of affairs, one should probably not assign a 
statistical confidence to nucleosynthesis results, to the extent that
these are still dominated by systematic, rather than statistical errors. 
Strictly speaking, what \fig{neffmg} really displays is the equivalent 
neutrino number $N_{eq}$ for various combinations of (\mnt,$g$) parameters 
that give the same helium abundance, rather than real limits. Of course, 
from these contours which contain the raw information an educated reader  
can judge which helium abundance should be considered plausible or not. 

We now illustrate in concrete models the fact that such
values of the \nt \nt Majoron coupling $\gsim 10^ {-4}$ 
are theoretically plausible.
Different models imply different expectations for the
Majoron coupling constants $g$ and the relation they
bear with the \nt mass $m_{\nu_\tau}$.
Our discussion so far is applicable to the simplest seesaw Majoron 
model of ref. \cite{CMP}. In this case one expects that \cite{774}
\beq
 g = \O \left( \frac{m_D^2}{M_R^2} \right)
\eeq
where ${m_D}$ is a typical Dirac neutrino mass and
${M_R} \propto \vev{\sigma}$ is the Majorana mass of the
right-handed \21 singlet neutrino. Clearly $g$ values in the
range required by nucleosynthesis are quite reasonable say,
for ${m_D} \sim 1 - 100$ GeV and ${M_R} \sim 10^4 - 10^8$ GeV.
Moreover, it is a good approximation in this model to neglect 
the s-channel scalar exchange diagram of \fig{fig2}.

There is a wide class of alternative Majoron models characterized 
by a low scale of lepton number violation \cite{mu,JoshipuraValle92,Zurab}.
These models are attractive because they lead to a wide variety of
processes which may be experimentally accessible \cite{beyond}. In 
this case one expects a simple direct correlation between the mass of 
the neutrinos and the magnitude of the diagonal couplings of neutrinos 
to Majorons. The \neu mass is simply the product of the Yukawa coupling
$g$ and the vacuum expectation value $\vev{\sigma}$ which 
characterizes the spontaneous violation of the global 
lepton number symmetry \cite{fae},
\beq
m = g \vev{\sigma}
\eeq
{}From this it follows that for \mnt $\sim$ 10 MeV and 
$\vev{\sigma} \sim$ 100 GeV one obtains $g \sim 10^{-4}$. 
This situation is therefore characteristic of models
where lepton number spontaneously breaks at the weak scale.

\begin{figure}
\centerline{
\psfig{file=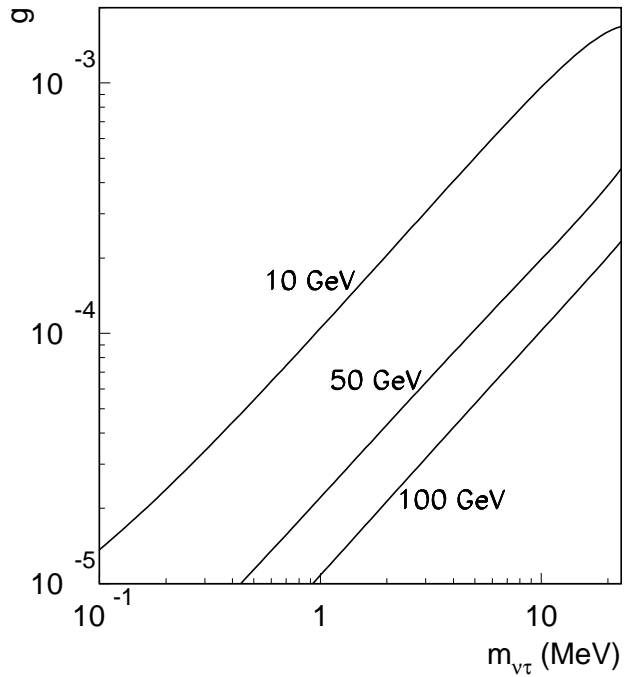,height=9cm}}
\caption{Expected values of $m_{\nu_\tau}$ and $g$ in model
of ref. [23]}
\label{neq3}
\end{figure}

There are more complicated models where the degree of correlation 
between the \nt mass $m_{\nu_\tau}$ and the lepton number violation 
scale may be different and may involve more free parameters.
Just to give a concrete example of such models, let us consider the 
supersymmetric models with spontaneous violation of
R parity \cite{MASIpot3}. These models lead to 
\beq
m \propto \frac{\vev{\sigma}^2}{M_{SUSY}}
\eeq
where $\vev{\sigma}$ is identified with the vacuum expectation
value of the right-handed \21 singlet sneutrino and  ${M_{SUSY}}$
denotes a typical neutralino mass. The expected values of
($g$,$m_{\nu_\tau}$) are depicted in \fig{neq3}, obtained when one
varies the other relevant free parameters over a theoretically 
reasonable range.

For all models with low-scale lepton number violation we have shown, 
by doing the full calculation, that the overall annihilation cross 
section for \nt \nt annihilation into two Majorons can be enhanced by 
an order of magnitude with respect to our above simplified calculation
which neglected the s-channel scalar exchange diagram in \fig{fig2}. 
Although this would allow us to weaken our limits, the effect on $g$ 
would only be a factor $ 10^{1/4} \lsim 2$, so that the limits 
derived in figure \ref{neffmg} could be relaxed by a factor $\lsim 2$
in this class of models.

As a last comment, we note that the limits obtained in our paper 
could also be tightened by including the  influence of non-equilibrium 
electronic neutrinos (and anti-neutrinos) produced by \nt \nt annihilations 
on the neutron-to-proton ratio \cite{noneq} but, again, the effect
is quite small on the bounds derived on $g$.

Last but not least, we must compare the limits obtained by primordial 
big bang nucleosynthesis with those derived from astrophysics.
A new light particle, like the Majoron, may have an important effect
on stellar evolution and this allows one to place stringent limits on 
the strength of the interaction of such particles \cite{Raffelt}. In the 
case we consider here, the Majorons interact predominantly with
a heavy $\nu_\tau$ (with the mass in MeV range), so its influence may
be noticeable in supernova explosions when the temperature reaches
tens of MeV. The bounds on Majoron properties which can be deduced from
supernova physics have been widely discussed \cite{arcadi} and 
recently analysed in ref. \cite{Raffelt} (see also references therein). 
For example a  Majoron with Yukawa 
coupling to electronic neutrinos in the range $10^{-6}-10^{-3}$ could 
be important for supernova physics. However in our model this coupling 
to \ne is much smaller. A Majoron coupling constant to tau-neutrinos around 
$10^{-4}$ may be potentially interesting for supernova physics and will 
be discussed elsewhere. Here we only mention that  $g$ values larger
than (a few)$\times 10^{-5} \sqrt {m/\mbox{MeV}}$ may be dangerous because 
the coupling is strong enough for abundant production of Majorons in 
high temperature regions in the supernova core and simultaneously 
small enough so that the mean free path of the produced Majorons is larger 
than the central stellar core. Still the coupling $g> 10^{-4}$ seems 
to be allowed. 

\section{Conclusions}

In this paper we have investigated the implications for primordial 
nucleosynthesis of a heavy tau neutrino in the MeV range, in the 
presence of sufficiently strong \nt annihilations into Majorons.
We have determined the effective neutrino number $N_{eq}$, or 
equivalently the primordial helium abundance, and studied the 
level of sensitivity that it exhibits when expressed in terms 
of the underlying \nt mass \mnt and coupling parameter $g$, the 
relevant coupling constant determining the \nt \nt annihilation 
cross section. Given the fact that present nucleosynthesis 
discussions are still plagued by systematics, it is useful to 
interpret our results this way, rather than as an actual limit 
in the statistical sense. 
For each \mnt value, one can in principle identify the corresponding
lower bounds on $g$ for which the \nt \nt annihilations to Majorons are 
sufficiently efficient in order not to be in conflict with nucleosynthesis.
Moreover, in contrast to the standard model, these models can
account for a value of  $N_{eq} \leq 3$ if the \nt mass lies in the
region 1-10 MeV, provided $\geq 3 \times 10^ {-4}$.

We have been conservative in determining the nucleosyhthesis
limits, to the extent that we have neglected model-dependent
contributions from s-channel Higgs boson exchange, given in
\fig{fig2}. This seems reasonable from the point of view of
the relevant particle physics models 
\cite{CMP,mu,JoshipuraValle92,Zurab,MASIpot3}.

We have also concluded that, indeed, the required choice of parameters 
can be naturally realized in Majoron models both with weak and large-scale 
lepton number violation. As a result, for sufficiently large but plausible
values of the \nt \nt Majoron coupling $\gsim 10^ {-4}$ one 
can not rule out any values of the \nt mass up the present 
laboratory limit based on the cosmological argument. This
highlights the importance of further experimental efforts
in laboratory searches for the \nt mass. Improvements expected 
at a tau-charm factory are indeed necessary, since the primordial 
nucleosynthesis constraints on the \nt mass can be easily relaxed 
in a large class of extensions of the standard electroweak model.

\section*{Appendix}

Here we show why one can neglect the s-channel diagram of \fig{fig2}
in the determination of the nucleosynthesis bound on $m_{\nu_\tau}$ and 
majoron coupling $g$. 

The total cross-section for the annihilation to Majorons that corresponds to
s-channel and t-channel diagrams of \fig{fig2} is given by

\begin{equation}
\label{sigmaJJ}
\sigma_J(\epsilon,\eta) = \frac{1}{64\pi} \frac{g^4}{m_{\nu_\tau}^2}
\left[ \epsilon^2 \sqrt{\eta} + (1 -2 \epsilon)
\frac{1-\eta}{2\eta}
\left[\ln \left(\frac{1+\sqrt{\eta}}{1-\sqrt{\eta}}\right) -
2\sqrt{\eta}\right] \right].
\end{equation}

\noi
where the parameter $\epsilon$ is defined by

\begin{equation}
\epsilon\equiv \frac{m_{\nu_\tau}}{g^2_{\nu \nu J}}
\sum_i \left( \frac{g_{\nu \nu H_i} g_{H_i J J}}{m^2_{H_i}} \right)
\label{epsdef}
\end{equation}

\noi
and $g_{\nu \nu H_i}$, $g_{H_i J J}$ are the couplings relevant for
the s-channel diagram of \fig{fig2}. In \Eq{epsdef} the sum is over
all the CP-even scalars present in the model. From its definition, 
one can see that $\epsilon$ is proportional to the couplings 
$\nu \nu H_i$ and $ H_i JJ$. When $\epsilon \ra 0$ the s-channel 
becomes zero. 

The value of $\epsilon$ depends very much on the model. For the
pure-singlet majoron models with low lepton number violation scale
considered in ref. \cite{JoshipuraValle92} there is a strict 
correlation between the \neu mass and the lepton number violation 
scale. In this case one has  $\epsilon =1$. For seesaw
models, with  lepton number violated at a large mass scale, 
one has  $\epsilon \ll 1$. For the supersymmetric model with 
spontaneous breaking of R parity \cite{MASIpot3} at the weak
scale one can show that $\epsilon$ typically lies in a range 
around the value 1/2. In our analysis we wanted to stay as 
much model independent as possible. In order to have an idea 
of the dependence of our results on $\epsilon$ we define
\begin{equation}
F(x,\epsilon) \equiv
\int_{0}^{1}d\eta~\frac{\eta}{(1-\eta)^{7/2}}
\sigma_J(\epsilon,\eta) K_1\Bigl (\frac{2x}{\sqrt{1-\eta}}\Bigr )
\end{equation}

\begin{figure}
\centerline{\psfig{file=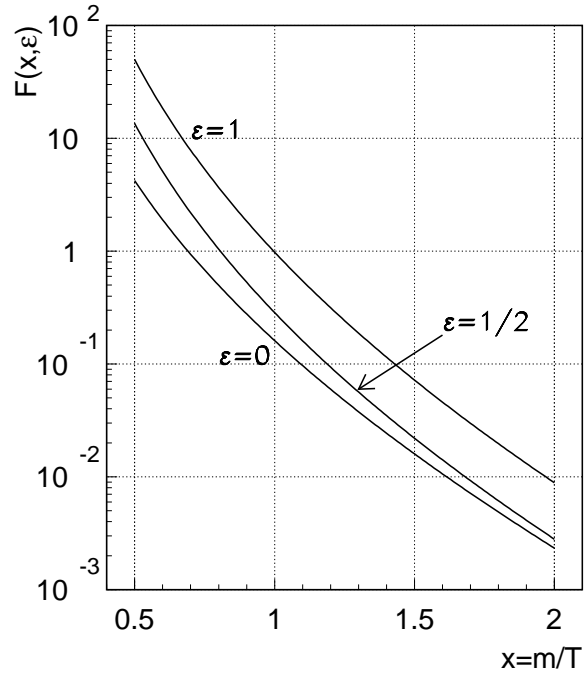,height=9cm}}
\caption{The function $F(x,\epsilon)$ for various values of $\epsilon$}
\label{fxeps}
\end{figure}

\noi 
which is just the integrand of \eq{intsigmav2} in section \ref{bef}. 
In \fig{fxeps} we plot the function $F(x,\epsilon)$ for $\epsilon=0,1/2$ 
and $1$. We see that the value $\epsilon=0$ represents a lower bound on 
that integral. For most models we would get a higher value. If we notice 
that the cross section is proportional to $g^4$, that diference in 
$F(x,\epsilon)$ would translate into a smaller value needed for $g$ 
to satisfy the nucleosynthesis bounds. Therefore, one can obtain 
a model-independent and conservative bound by taking the worst possible
case, which corresponds to $\epsilon=0$. Due to the dependence of $F$ 
on $g^4$ the bounds on $g$ would not be too sensitive to the value of
$\epsilon$ in the range of interest. This justifies our simplified 
expression for $\sigma_J$ used in \eq{sigmaJ}.

\section*{Acknowledgements}

This work has been supported by DGICYT under Grants PB95-1077
and SAB94-0089 (A. D.), by the TMR network grant ERBFMRXCT960090 
of the European Union, and by an Acci\'on Integrada Hispano-Portuguesa.
S.P. was supported by Conselleria d'Educaci\'o i Ci\`encia of Generalitat 
Valenciana. A.D. also acknowledges the support of the Danish National
Science Research Council through grant 11-9640-1 and in part
of the Danmarks Grundforskningsfond through its
support of the Theoretical Astrophysical Center. We thank
A. Santamaria for fruitful discussions.

\end{document}